\documentclass[conference]{IEEEtran}
\usepackage{balance}
\usepackage{cite}
\usepackage{amsmath,amssymb,amsfonts}
\usepackage{algorithmic}
\usepackage{graphicx}
\usepackage{textcomp}
\usepackage{xcolor}
\usepackage{url}
\usepackage{hyperref}
\usepackage[caption=false,font=footnotesize]{subfig}
\usepackage{booktabs}
\usepackage{tikz}
\usetikzlibrary{arrows.meta, calc, decorations.pathreplacing, positioning}
\usepackage[linesnumbered, ruled, vlined]{algorithm2e}
\DontPrintSemicolon

\SetCommentSty{AlgoCommentStyle}
\SetKwComment{tcp}{//~}{}
\SetKwProg{Fn}{Function}{}{}

\newif\ifreviewcopy
\ifdefined\reviewcopyenabled
  \reviewcopytrue
\else
  \reviewcopyfalse
\fi

\newcommand{\papertitle}{cuSBF: A Minimizer-Aware Bloom Filter for Genomic Sequence Data on Modern GPUs}
\newcommand{\paperkeywords}{Bloom filters, GPU computing, approximate membership query, minimizers, genomic indexing}
\newcommand{\publicrepourl}{https://github.com/tdortman/cuSBF}

\ifreviewcopy
\newcommand{\repourl}{\anonymousrepourl}
\newcommand{\paperauthormeta}{Anonymous Authors}
\newcommand{\reviewcopymeta}{true}
\newcommand{\paperauthors}{
    \author{
        \IEEEauthorblockN{Anonymous Authors}
    }
}
\else
\newcommand{\repourl}{\publicrepourl}
\newcommand{\paperauthormeta}{Tim Dortmann, Markus Vieth, Bertil Schmidt}
\newcommand{\reviewcopymeta}{false}
\newcommand{\paperauthors}{
    \author{
        \IEEEauthorblockN{Tim Dortmann}
        \IEEEauthorblockA{
            \emph{Institute of Computer Science} \\
            \emph{Johannes Gutenberg University}\\
            Mainz, Germany \\
            research@tdortman.com
        }
        \and
        \IEEEauthorblockN{Markus Vieth}
        \IEEEauthorblockA{
            \emph{Institute of Computer Science} \\
            \emph{Johannes Gutenberg University}\\
            Mainz, Germany \\
            vieth@uni-mainz.de
        }
        \and
        \IEEEauthorblockN{Bertil Schmidt}
        \IEEEauthorblockA{
            \emph{Institute of Computer Science} \\
            \emph{Johannes Gutenberg University}\\
            Mainz, Germany \\
            bertil.schmidt@uni-mainz.de
        }
    }
}
\fi

\begin{document}

\title{\papertitle}
\hypersetup{
    pdftitle={\papertitle},
    pdfauthor={\paperauthormeta},
    pdfkeywords={\paperkeywords},
    pdfinfo={
        RepositoryURL={\repourl},
        ReviewCopy={\reviewcopymeta}
    }
}

\paperauthors
\maketitle

\begin{abstract}
Efficient genomic $k$‑mer indexing depends on approximate membership query (AMQ) structures that must deliver high throughput, low false‑positive rates (FPR), and modest memory footprints. The Super Bloom filter (SBF) is attractive for this scenario because minimizer‑guided sharding and the Findere scheme exploit the redundancy of overlapping $k$‑mers. However, those same features cause high per‑$k$‑mer compute cost, severe register pressure, and irregular memory accesses, which hinder an effective GPU implementation. We present \emph{cuSBF}, an open‑source, header‑only CUDA library that implements SBF for sequence‑native workloads. cuSBF’s design merges sectorized shards, cooperative shared‑memory tiling, warp‑level shard sharing, and segmented warp reductions, turning super‑$k$‑mer locality into scalable GPU parallelism.
%Genomic $k$-mer indexing relies heavily on approximate membership query structures, which must deliver high throughput and low false-positive rates under tight memory budgets. The Super Bloom filter is attractive for this setting because minimizer-guided sharding and the findere scheme exploit sequence overlaps, but these same features make an efficient GPU implementation challenging due to high per-$k$-mer compute cost, register pressure, and irregular memory accesses. We present \emph{cuSBF}, an open-source, header-only CUDA implementation of the Super Bloom filter for sequence-native workloads. Its design combines sectorized shards, cooperative shared-memory tiling, warp-level shard sharing, and segmented warp reductions to turn super-$k$-mer locality into effective GPU parallelism.
Across real genomic workloads on RTX PRO 6000 Blackwell and GH200 systems, cuSBF achieves the highest throughput among all evaluated sequence-capable baselines.
On the RTX PRO 6000, it surpasses the \mbox{cuCollections} blocked Bloom filter baseline by up to 9.1$\times$ for insertion and 7.7$\times$ for query, while reaching up to 92$\times$ and 234$\times$ speedups over the multi-threaded CPU Super Bloom reference implementation. It also outperforms GPU-based dynamic AMQs (Cuckoo, Two-Choice, Quotient filters) by $1.5$–$3400\times$ depending on workload characteristics.
A parameter sweep identifies $(s=28, m=16, H=4)$ as Pareto-optimal for $k=31$, yielding significantly lower FPR than cuCollections at matched memory budgets.
%A comprehensive parameter sweep over substring size for the Findere scheme ($s$), minimizer size ($m$), and number of hash functions ($H$) identifies the configuration $(s=28,\;m=16,\;H=4)$ as optimal for $k=31$.
%A parameter sweep over $(s,m,H)$ identifies $s=28$, $m=16$, and $H=4$ as the best overall operating point for $k=31$.
%At comparable memory budgets, this configuration also yields consistently lower false-positive rates than the GPU blocked Bloom baseline.
%Our results show that minimizer-aware approximate membership queries can be mapped efficiently to modern GPUs and can substantially outperform generic GPU filters on genomic sequence data.
Crucially, cuSBF’s architecture-aware design sustains 85\% streaming multiprocessor utilization even for out-of-cache filters — proving that sequence locality, not raw bandwidth, is the key to GPU-accelerated genomic indexing.

\end{abstract}


\section{Introduction}

Approximate membership query (AMQ) data structures answer the question "is element $q$ in set $S$?" while allowing a tunable false‑positive rate (FPR) $\epsilon$, thereby trading a small amount of inaccuracy for substantial savings in memory and time. The Bloom filter  \cite{bloom}  has been the de‑facto AMQ structure for more than two decades and is employed in a wide range of domains, from databases \cite{chang2008bigtable} and networking \cite{bloom-network} to bioinformatics \cite{gaia2019ngsreadstreatment}.
The Blocked Bloom filter  \cite{blocked-bloom} improves cache locality by partitioning the bit array into small blocks; GPU‑accelerated blocked Bloom filter implementations \cite{cuCollections,jünger2025optimizingbloomfiltersmodern}  have demonstrated that such structures can fully saturate the global‑memory bandwidth of modern GPUs.

%Approximate membership query (AMQ) data structures answer the question "is element $x$ in set $S$?" with a small, controllable false positive rate (FPR) $\epsilon$, trading accuracy for significant gains in space and time efficiency. For decades, the Bloom filter \cite{bloom} has been the standard AMQ solution, seeing widespread use anywhere from databases \cite{chang2008bigtable} and networking \cite{bloom-network} to bioinformatics \cite{gaia2019ngsreadstreatment}. Its blocked variant \cite{blocked-bloom} improved cache locality by partitioning the bit array into small blocks, and GPU-accelerated blocked Bloom filters \cite{cuCollections,jünger2025optimizingbloomfiltersmodern} have demonstrated that they can effectively saturate GPU global memory bandwidth.

For genomic sequence analysis, and in particular $k$‑mer indexing, the Super Bloom filter (SBF) \cite{superbloom} adopts a fundamentally different strategy. Instead of hashing each $k$‑mer independently, SBF performs a two‑level decomposition: a minimizer hash selects a Bloom‑filter shard, and a set of $s$‑mer hashes sets bits inside that shard. Because consecutive genomic $k$‑mers often share the same minimizer, SBF exploits this intrinsic locality to reduce the number of distinct insertions and queries, yielding higher throughput and lower FPR on CPUs. Nevertheless, CPU‑bound performance remains limited by memory‑bandwidth and compute constraints.

%For sequence data, particularly genomic $k$-mer indexing, the Super Bloom filter (SBF) \cite{superbloom} introduces a fundamentally different approach. Rather than hashing each $k$-mer independently as a random integer, SBF decomposes the problem into two levels: a minimizer hash selects a Bloom filter block (shard), and $s$-mer hashes set bits within that shard. This structure exploits the sequence locality inherent in consecutive $k$-mers, which very often share minimizers. On CPUs, SBF offers competitive throughput at much lower false positive rates, but its throughput remains limited by the CPU's memory bandwidth and compute.

Porting SBF to GPUs introduces three major challenges.

\begin{itemize}
    \item \textbf{High per‑$k$‑mer compute cost.} A naïve GPU implementation would (i) pack a $k$‑mer into a 64‑bit word, (ii) slide a window of length $(k-m+1)$ to locate the minimizer, (iii) hash $(k-s+1)$ $s$‑mers, and (iv) apply $H$ hash functions to set bits in the selected shard. This approach can dominate the ALU pipeline, causing instruction stalls that leave the memory subsystem underutilized.

    \item \textbf{Severe register pressure.}  All intermediate values (packed $k$‑mer, minimizer hash, per‑$s$‑mer masks, shard index, and tile pointers) must reside in registers for latency hiding. The resulting high register usage limits the number of concurrent warps per Streaming Multiprocessor (SM), reducing occupancy and starving the memory subsystem of in‑flight requests.

    \item \textbf{Irregular memory access patterns.} While many adjacent $k$‑mers share a minimizer (and thus the same shard), minimizer boundaries cause threads within a warp to diverge and scatter to different shards. Without coordinated warp‑level handling this leads to redundant shard loads, wasteful global‑memory traffic, and destructive atomic contention during insertions.
\end{itemize}

%Porting SBF to GPUs, however, presents unique challenges. Unlike a standard Bloom filter where each element maps to $H$ independent bit positions, SBF must perform substantially more per-$k$-mer work: packing $k$ symbols into a 64-bit integer, scanning $(k-m+1)$ $m$-mer windows for the minimizer, hashing $(k-s+1)$ $s$-mer windows, and distributing the resulting bits across $H$ hash functions. This compute-intensive work can easily overwhelm the GPU's ALU pipeline, causing instruction stalls that leave memory bandwidth underutilized. Furthermore, keeping intermediate values (packed $k$-mer, minimizer hash, four word masks, shard index, and tile pointers) entirety in registers is essential for latency hiding, but this high register pressure limits the number of concurrent warps per SM. A na\"ive implementation can easily end up with very low occupancy, starving the memory subsystem of in-flight requests. A third challenge is that the minimizer-guided sharding pattern produces irregular memory access: consecutive $k$-mers often share a minimizer (targeting the same shard), but minimizer boundaries still introduce disjoint execution regions where adjacent threads scatter to different shards. Without careful warp-level coordination, this pattern wastes global memory bandwidth on redundant shard loads and causes destructive atomic contention during insertion.

In this paper we introduce \emph{cuSBF}\footnote{cuSBF is released as open‑source software (header‑only CUDA) at: \\ \url{\repourl}}, a high‑performance, header‑only CUDA library that implements the Super Bloom filter for sequence‑native workloads on modern GPUs. Our key insight is that the compute overhead and register pressure can be mitigated by amortizing work across multiple $k$‑mers per thread and by using simplified hash functions, while the minimizer‑driven locality can be turned into a performance advantage through warp‑level coordination.
Our design combines a sectorized 256‑bit shard layout, cooperative shared‑memory tiling, warp‑level shard sharing, and segmented warp reductions, thereby converting super‑$k$‑mer locality into effective GPU parallelism.

%In this paper, we present \emph{cuSBF}, a high-performance, header-only CUDA library that accelerates the Super Bloom filter on modern GPUs. \footnote{The source code is available as an open-source header-only library at: \url{\repourl}} Our key insight is that the compute overhead and register pressure can be tamed through careful amortization (processing multiple $k$-mers per thread) and simplified hash functions, while the minimizer-induced latency can be exploited at the warp level to coalesce memory accesses and reduce atomic operations.

We make the following contributions:

\begin{itemize}
    \item \textbf{High-performance CUDA SBF library}: We present a sequence-native, header-only implementation that supports host and device sequences, dense record batches, and streamed
    FASTA/FASTQ input, including gzip-compressed files.

    \item \textbf{Novel GPU-centric SBF design}: We introduce a sectorized 256-bit shard layout, cooperative shared-memory sequence tiling, warp-level shard sharing, and segmented warp reductions.

    \item \textbf{Comprehensive Evaluation}: Throughput and FPR analysis on RTX PRO 6000 (Blackwell) and GH200 GPUs, a full parameter sweep over $(s,m,H)$  that reveals Pareto‑optimal configurations, and a speed‑of‑light analysis distinguishing compute‑bound from bandwidth‑bound regimes.
    %Throughput and FPR analysis across GPU and CPU baselines, a parameter sweep over $(s,m,H)$ revealing Pareto-optimal configurations, and a speed-of-light analysis contrasting compute-bound and bandwidth-bound behavior.
\end{itemize}

The rest of the paper is organized as follows.
Section~\ref{sec:background} surveys Bloom filter variants (standard, blocked, and Super Bloom) and introduces the GPU architectural features that drive our design. 
Section~\ref{sec:related-work} reviews prior GPU‑based AMQ structures and minimizer‑driven sequence‑indexing techniques.
Section~\ref{sec:design}  details the cuSBF library, covering the sectorized‑shard layout, cooperative shared‑memory tiling, warp‑level insertion and query algorithms, and the sequence‑native FASTX ingestion pipeline.
Section~\ref{sec:evaluation} presents the experimental methodology, the $(s,m,H)$  parameter sweep, throughput and false‑positive evaluations, a speed‑of‑light analysis, and a study of host‑to‑device transfer overhead.
Section~\ref{sec:conclusion} concludes the paper.
%Section~\ref{sec:background} reviews standard, blocked, and Super Bloom filters together with info on the relevant GPU architectures, and Section~\ref{sec:related-work} surveys related GPU AMQs and minimizer-based sequence indexing. Section~\ref{sec:design} presents the cuSBF library design, including sectorized shards, cooperative shared-memory tiling, warp-level insertion and query algorithms, and sequence-native FASTX ingestion. Section~\ref{sec:evaluation} describes our experimental setup, the $(s,m,H)$ parameter sweep, throughput and false-positive comparisons, a speed-of-light analysis, and host-to-device transfer overhead. Section~\ref{sec:conclusion} concludes.

\section{Background}
\label{sec:background}

\subsection{Bloom Filters}

A Bloom filter \cite{bloom} represents a set $S$ using an array of $m$ bits, initially all zero.
Upon insertion of an element $x$, $H$ independent hash functions map $x$ to $H$ bit positions, which are set to 1.
A query for $q$ checks whether all $H$ positions are set: if any is 0, $q \notin S$ definitely (See Fig.~\ref{fig:bloom-filter-std}).
If all are 1, $q \in S$ with probability $1-\epsilon$, where the FPR for $n$ inserted elements is approximately

\begin{equation}
    \epsilon \approx \left(1 - e^{-Hn/m}\right)^H.
\end{equation}

The \emph{Blocked Bloom filter} (BBF) \cite{blocked-bloom} partitions the $m$ bits into $B$ equal-sized blocks (shards).
Each element is assigned to exactly one block, and bits are set only within that block (Fig.~\ref{fig:bloom-filter-blocked}).
While this does improve cache locality on CPUs and GPUs, isolated blocks cannot "average" collisions across the full array, leading to higher FPR at a given memory budget.

\definecolor{insertRed}{RGB}{205,68,52}
\definecolor{insertGreen}{RGB}{55,148,84}
\definecolor{insertBlue}{RGB}{66,112,214}
\definecolor{queryGray}{RGB}{92,92,92}
\definecolor{panelFill}{RGB}{248,249,252}
\definecolor{panelStroke}{RGB}{212,217,226}
\definecolor{bitOnFill}{RGB}{229,238,251}
\definecolor{bitOnDraw}{RGB}{90,126,192}
\definecolor{missRed}{RGB}{186,76,76}
\definecolor{textDark}{RGB}{38,38,38}

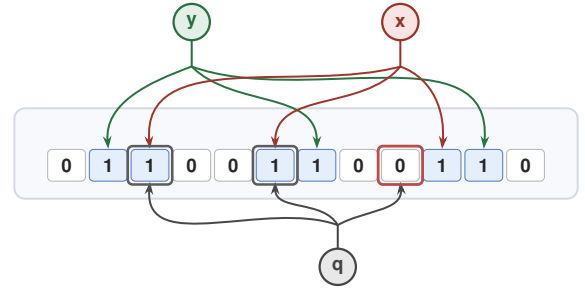
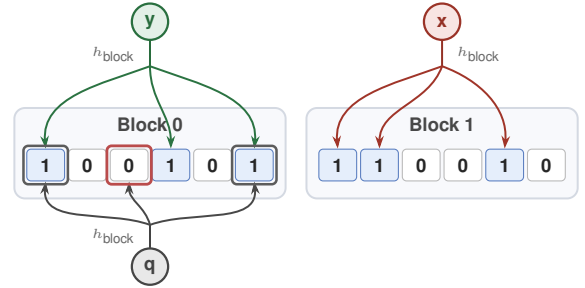
\begin{figure}[t]
  \centering
  \subfloat[Standard Bloom filter ($m = 12$, $k = 3$).\label{fig:bloom-filter-std}]{%
    \begin{minipage}[b]{0.48\textwidth}
      \centering
      \begin{tikzpicture}[
  x=0.92cm,
  y=0.92cm,
  font=\sffamily\scriptsize,
  >=Stealth,
  every node/.style={text=textDark},
  panel/.style={
    rounded corners=4pt,
    draw=panelStroke,
    fill=panelFill,
    line width=0.75pt
  },
  bit/.style={
    rounded corners=1.6pt,
    draw=black!28,
    fill=white,
    minimum width=0.50cm,
    minimum height=0.42cm,
    inner sep=0pt,
    font=\sffamily\scriptsize\bfseries
  },
  biton/.style={
    bit,
    draw=bitOnDraw,
    fill=bitOnFill
  },
  item/.style={
    circle,
    minimum size=0.48cm,
    inner sep=0pt,
    font=\sffamily\scriptsize\bfseries,
    draw=#1!75!black,
    fill=#1!14,
    text=#1!72!black,
    line width=0.85pt
  },
  item/.default=black,
  pathline/.style={draw=#1!72!black, line width=0.75pt, line cap=round, line join=round},
  pathline/.default=black,
  arrow/.style={-{Stealth[length=1.7mm,width=1.15mm]}, draw=#1!78!black, line width=0.75pt},
  arrow/.default=black,
  title/.style={font=\sffamily\scriptsize\bfseries, text=black!72},
  small/.style={font=\sffamily\tiny, text=black!60},
  hitoutline/.style={rounded corners=1.8pt, draw=queryGray, line width=0.95pt},
  missoutline/.style={rounded corners=1.8pt, draw=missRed, line width=1.05pt}
]

  % Panel.
  \path[panel] (-0.15,-0.36) rectangle (7.95,0.94);

  % Bit cells.
  \node[bit]   (b0)  at (0.6,0.12) {0};
  \node[biton] (b1)  at (1.2,0.12) {1};
  \node[biton] (b2)  at (1.8,0.12) {1};
  \node[bit]   (b3)  at (2.4,0.12) {0};
  \node[bit]   (b4)  at (3.0,0.12) {0};
  \node[biton] (b5)  at (3.6,0.12) {1};
  \node[biton] (b6)  at (4.2,0.12) {1};
  \node[bit]   (b7)  at (4.8,0.12) {0};
  \node[bit]   (b8)  at (5.4,0.12) {0};
  \node[biton] (b9)  at (6.0,0.12) {1};
  \node[biton] (b10) at (6.6,0.12) {1};
  \node[bit]   (b11) at (7.2,0.12) {0};

  % Inserted items.
  \node[item=insertGreen] (y) at (2.4,2.18) {y};
  \node[item=insertRed]   (x) at (5.4,2.18) {x};

  % Query item.
  \node[item=queryGray] (q) at (4.5,-1.34) {q};

  % Green insertion.
  \coordinate (yg) at (2.4,1.54);
  \draw[pathline=insertGreen] (y.south) -- (yg);
  \draw[arrow=insertGreen] (yg) to[out=-160,in=95] (b1.north);
  \draw[arrow=insertGreen] (yg) to[out=-45,in=90] (b6.north);
  \draw[arrow=insertGreen] (yg) to[out=-20,in=85] (b10.north);

  % Red insertion.
  \coordinate (xr) at (5.4,1.54);
  \draw[pathline=insertRed] (x.south) -- (xr);
  \draw[arrow=insertRed] (xr) to[out=-165,in=95] (b2.north);
  \draw[arrow=insertRed] (xr) to[out=-130,in=90] (b5.north);
  \draw[arrow=insertRed] (xr) to[out=-20,in=85] (b9.north);

  % Query from below (2 hits: b2 and b5, 1 miss: b8).
  \coordinate (qq) at (4.5,-0.74);
  \draw[pathline=queryGray] (q.north) -- (qq);
  \draw[arrow=queryGray] (qq) to[out=160,in=-95] (b2.south);
  \draw[arrow=queryGray] (qq) to[out=120,in=-90] (b5.south);
  \draw[arrow=queryGray] (qq) to[out=30,in=-85]  (b8.south);

  % Query result highlighting.
  \draw[missoutline] ($(b8.south west)+(-0.04,-0.04)$) rectangle ($(b8.north east)+(0.04,0.04)$);
  \draw[hitoutline]  ($(b2.south west)+(-0.04,-0.04)$) rectangle ($(b2.north east)+(0.04,0.04)$);
  \draw[hitoutline]  ($(b5.south west)+(-0.04,-0.04)$) rectangle ($(b5.north east)+(0.04,0.04)$);

\end{tikzpicture}
    \end{minipage}%
  }
  \hfill
  \subfloat[Blocked Bloom filter ($m = 12$, $B = 2$).\label{fig:bloom-filter-blocked}]{%
    \begin{minipage}[b]{0.48\textwidth}
      \centering
      \begin{tikzpicture}[
    x=0.92cm,
    y=0.92cm,
    font=\sffamily\scriptsize,
    >=Stealth,
    every node/.style={text=textDark},
    panel/.style={
      rounded corners=4pt,
      draw=panelStroke,
      fill=panelFill,
      line width=0.75pt
    },
    bit/.style={
      rounded corners=1.6pt,
      draw=black!28,
      fill=white,
      minimum width=0.50cm,
      minimum height=0.42cm,
      inner sep=0pt,
      font=\sffamily\scriptsize\bfseries
    },
    biton/.style={
      bit,
      draw=bitOnDraw,
      fill=bitOnFill
    },
    item/.style={
      circle,
      minimum size=0.48cm,
      inner sep=0pt,
      font=\sffamily\scriptsize\bfseries,
      draw=#1!75!black,
      fill=#1!14,
      text=#1!72!black,
      line width=0.85pt
    },
    item/.default=black,
    pathline/.style={draw=#1!72!black, line width=0.75pt, line cap=round, line join=round},
    pathline/.default=black,
    arrow/.style={-{Stealth[length=1.7mm,width=1.15mm]}, draw=#1!78!black, line width=0.75pt},
    arrow/.default=black,
    title/.style={font=\sffamily\scriptsize\bfseries, text=black!72},
    small/.style={font=\sffamily\tiny, text=black!60},
    hitoutline/.style={rounded corners=1.8pt, draw=queryGray, line width=0.95pt},
    missoutline/.style={rounded corners=1.8pt, draw=missRed, line width=1.05pt}
  ]

  % Panels and headings.
  \path[panel] (-0.15,-0.36) rectangle (3.75,0.94);
  \path[panel] (4.05,-0.36) rectangle (7.95,0.94);
  \node[title, fill=panelFill, inner sep=1.5pt] at (1.80,0.70) {Block 0};
  \node[title, fill=panelFill, inner sep=1.5pt] at (6.00,0.70) {Block 1};

  % Bit cells (Block 0).
  \node[biton] (b0_0) at (0.3,0.12) {1};
  \node[bit]   (b0_1) at (0.9,0.12) {0};
  \node[bit]   (b0_2) at (1.5,0.12) {0};
  \node[biton] (b0_3) at (2.1,0.12) {1};
  \node[bit]   (b0_4) at (2.7,0.12) {0};
  \node[biton] (b0_5) at (3.3,0.12) {1};

  % Bit cells (Block 1).
  \node[biton] (b1_0) at (4.5,0.12) {1};
  \node[biton] (b1_1) at (5.1,0.12) {1};
  \node[bit]   (b1_2) at (5.7,0.12) {0};
  \node[bit]   (b1_3) at (6.3,0.12) {0};
  \node[biton] (b1_4) at (6.9,0.12) {1};
  \node[bit]   (b1_5) at (7.5,0.12) {0};

  % Inserted items.
  \node[item=insertGreen] (y) at (1.80,2.18) {y};
  \node[item=insertRed]   (x) at (6.00,2.18) {x};

  % Query item.
  \node[item=queryGray] (q) at (1.80,-1.34) {q};

  % Green insertion.
  \coordinate (yg) at (1.80,1.54);
  \draw[pathline=insertGreen] (y.south) -- (yg) node[midway, left=2pt, small] {$h_{\text{block}}$};
  \draw[arrow=insertGreen] (yg) to[out=-160,in=95] (b0_0.north);
  \draw[arrow=insertGreen] (yg) to[out=-90,in=90] (b0_3.north);
  \draw[arrow=insertGreen] (yg) to[out=-20,in=85] (b0_5.north);

  % Red insertion.
  \coordinate (xr) at (6.00,1.54);
  \draw[pathline=insertRed] (x.south) -- (xr) node[midway, right=2pt, small] {$h_{\text{block}}$};
  \draw[arrow=insertRed] (xr) to[out=-160,in=95] (b1_0.north);
  \draw[arrow=insertRed] (xr) to[out=-110,in=90] (b1_1.north);
  \draw[arrow=insertRed] (xr) to[out=-20,in=85] (b1_4.north);

  % Query from below (2 hits: b0_0 and b0_5, 1 miss: b0_2).
  \coordinate (qq) at (1.80,-0.74);
  \draw[pathline=queryGray] (q.north) -- (qq) node[midway, left=2pt, small] {$h_{\text{block}}$};
  \draw[arrow=queryGray] (qq) to[out=160,in=-95] (b0_0.south);
  \draw[arrow=queryGray] (qq) to[out=105,in=-90] (b0_2.south);
  \draw[arrow=queryGray] (qq) to[out=20,in=-85]  (b0_5.south);

  % Query result highlighting.
  \draw[missoutline] ($(b0_2.south west)+(-0.04,-0.04)$) rectangle ($(b0_2.north east)+(0.04,0.04)$);
  \draw[hitoutline]  ($(b0_0.south west)+(-0.04,-0.04)$) rectangle ($(b0_0.north east)+(0.04,0.04)$);
  \draw[hitoutline]  ($(b0_5.south west)+(-0.04,-0.04)$) rectangle ($(b0_5.north east)+(0.04,0.04)$);

\end{tikzpicture}
    \end{minipage}%
  }
  \caption{Comparison of Bloom filter variants.
  In the standard Bloom filter (a), $k = 3$ hash functions map each item globally across all $m = 12$ bits.
  In the blocked Bloom filter (b), a block-selection hash function ($h_{\text{block}}$) first maps each item to one of the $B = 2$ blocks, and then $k = 3$ in-block hash functions map the item within that chosen block.
  A query for $q$ returns ``no'' because one of its target bits is $0$ (despite two positive hits).}
  \label{fig:bloom-filter-comparison}
  \label{fig:blocked-bloom-filter}
\end{figure}

\subsection{Super Bloom Filter}

The \emph{Super Bloom filter} (SBF)~\cite{superbloom} adapts blocked Bloom filters to streaming sequence data such as genomic strings by exploiting two structural properties of consecutive $k$-mers: they overlap heavily, and they can be grouped by shared minimizers.

\subsubsection{\textbf{Minimizers and Super-k-mers}}
Consider a sequence $S$ over an alphabet $\Sigma$. A \(k\)-mer of $S$ is any substring of length $k$ contained within $S$.
Each $k$-mer contains $w = k-m+1$ overlapping substrings of length $m$, called $m$-mers.
Given a hash function over all $m$-mers, the \emph{minimizer} \cite{minimizers} of a $k$-mer is the $m$-mer with the minimum hash value.
Consecutive $k$-mers overlap by $k-1$ symbols, and therefore by $w-1$ of their $m$-mers.
As a result, adjacent $k$-mers frequently select the same minimizer. Fig. \ref{fig:minimizer-sharding} illustrates an example for $S = ACTGAAACTTAG$ over $\Sigma = \{A, C, G, T\}$ and $k = 9,\;m = 5$.

A {\bf super-$k$-mer} is a maximal run of consecutive $k$-mers sharing the same minimizer.
The density $d$ of a minimizer scheme is the fraction of $m$-mers that become minimizers.
For random minimizers on random sequences, the expected density is well-approximated by $d \approx 2/(w+1)$ \cite{superbloom}, so the expected super-$k$-mer length is $1/d \approx (w+1)/2$.

For our default cuSBF configuration $k = 31, \;m=16$, we have $w = 31-16+1 = 16$, giving an expected super-$k$-mer length of $(16+1)/2 = 8.5$.
This means that, on average, about 8 consecutive $k$-mers map to the same shard.
SBF assigns all $k$-mers in a super-$k$-mer to the shard selected by their shared minimizer, amortizing the shard load across the entire run (Fig. \ref{fig:minimizer-sharding}).

\subsubsection{\textbf{Findere Scheme}}

Within the selected shard, SBF does not insert or query $k$-mers directly. Instead, it uses the Findere scheme \cite{findere}: for a $k$-mer, all $z+1 = k-s+1$ overlapping $s$-mers (with $s<k$) are inserted or tested.
A query returns positive only if every one of the $z+1$ $s$-mers is found in the shard.

This approach exponentially suppresses false positives relative to the base $s$-mer false-positive rate.
For a random alien $k$-mer that shares no $s$-mers with indexed data, each of its $z+1$ $s$-mers must independently produce a false positive.
If a single $s$-mer false-positive rate is $\epsilon$, the $k$-mer false-positive rate is approximately

\begin{equation}
    \epsilon_{k\text{-mer}} \approx \epsilon^{\,z+1} = \epsilon^{\,k-s+1}.
    \label{eq:findere-fpr}
\end{equation}

With the default cuSBF configuration $k=31,\;s=28$, each $k$-mer is represented by $z+1=4$
overlapping $s$-mers.
With $H=4$ hash functions per $s$-mer, a random alien $k$-mer must satisfy $4 \times 4 = 16$ independent bit-tests simultaneously to yield a false positive.
Conversely, an \emph{almost} alien $k$-mer that differs from an indexed $k$-mer by only one base shares $z$ of its $z+1$ $s$-mers with that indexed $k$-mer, so only one alien $s$-mer must be falsely recognized.
Its FPR then reverts to approximately $\epsilon$ instead of $\epsilon^{\,z+1}$.
The parameter $s$ therefore tunes a trade-off: smaller $s$ increases $z+1$, strengthening the suppression of random aliens, while larger $s$ make each $s$-mer more selective, so a single-base mutation invalidates a larger fraction of all $s$-mers and near-matches are thus more strongly differentiated from true positives.

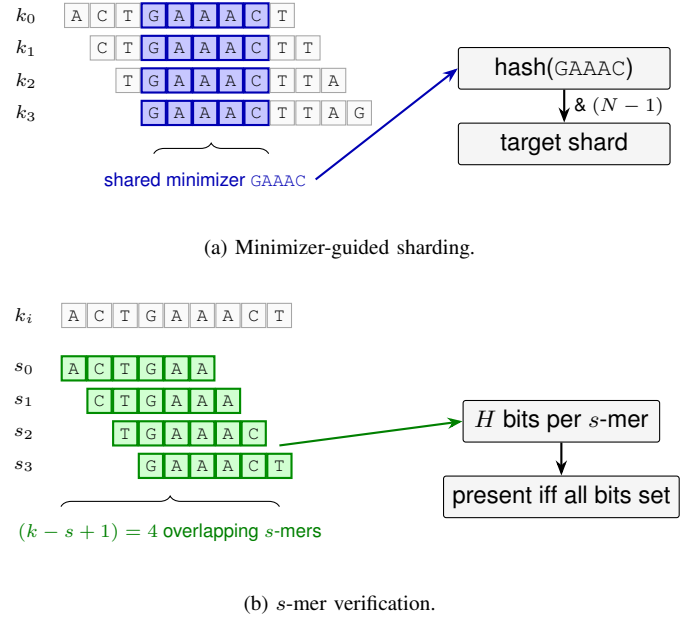
\begin{figure}[t]
\centering
\subfloat[Minimizer-guided sharding.\label{fig:minimizer-sharding}]{%
    \begin{minipage}[t]{\columnwidth}
        \centering
        \begin{tikzpicture}[
    font=\sffamily\small,
    base/.style={
        rectangle, 
        draw=black!35, 
        minimum width=0.32cm, 
        minimum height=0.32cm, 
        inner sep=0pt, 
        font=\ttfamily\scriptsize
    },
    hl/.style={
        base, 
        fill=blue!22, 
        draw=blue!70!black, 
        thick
    },
    box/.style={
        rectangle, 
        draw=black, 
        rounded corners=1pt, 
        fill=black!4, 
        minimum height=0.55cm, 
        align=center
    },
    arrow/.style={
        -{Stealth[length=2mm]}, 
        thick
    },
    note/.style={
        font=\sffamily\scriptsize
    }
  ]

  \def\seq{{"A","C","T","G","A","A","A","C","T","T","A","G","C","T"}}
  \foreach \row/\start in {0/0,1/1,2/2,3/3} {
    \node[note, anchor=east] at (-4.5,-0.25-0.43*\row) {$k_{\row}$};
    \foreach \i in {0,...,8} {
      \pgfmathtruncatemacro{\idx}{\start+\i}
      \pgfmathtruncatemacro{\isMin}{(\idx>=3) && (\idx<=7)}
      \pgfmathsetmacro{\x}{-4.1 + 0.34*(\start+\i)}
      \pgfmathsetmacro{\y}{-0.25 - 0.43*\row}
      \ifnum\isMin=1
      \node[hl] at (\x,\y) {\pgfmathparse{\seq[\idx]}\pgfmathresult};
      \else
      \node[base, fill=black!2] at (\x,\y) {\pgfmathparse{\seq[\idx]}\pgfmathresult};
      \fi
    }
  }

  \draw[decorate,decoration={brace,amplitude=4pt}] (-3.08,-2.08) -- (-1.56,-2.08);
  \node[note, blue!70!black] (minlabel) at (-2.40,-2.42) {shared minimizer \texttt{GAAAC}};

  \node[box, minimum width=2.8cm] (minhash) at (2.35,-0.95) {hash(\texttt{GAAAC})};
  \node[box, minimum width=2.8cm] (shard) at (2.35,-1.95) {target shard};
  \draw[arrow, blue!70!black] (minlabel.east) -- (minhash.west);
  \draw[arrow] (minhash.south) -- node[right,note] {\& $(N-1)$} (shard.north);

\end{tikzpicture}
    \end{minipage}%
}

\vspace{0.4em}

\subfloat[$s$-mer verification.\label{fig:findere-scheme}]{%
    \begin{minipage}[t]{\columnwidth}
        \centering
        \begin{tikzpicture}[
    font=\sffamily\small,
    base/.style={
        rectangle,
        draw=black!35,
        minimum width=0.32cm,
        minimum height=0.32cm,
        inner sep=0pt,
        font=\ttfamily\scriptsize
    },
    hl/.style={
        base,
        fill=green!18,
        draw=green!55!black,
        thick
    },
    box/.style={
        rectangle,
        draw=black,
        rounded corners=1pt,
        fill=black!4,
        minimum height=0.55cm,
        align=center
    },
    arrow/.style={
        -{Stealth[length=2mm]},
        thick
    },
    note/.style={
        font=\sffamily\scriptsize
    }
]

\def\seq{{"A","C","T","G","A","A","A","C","T"}}

\node[note, anchor=east] at (-4.5,0.25) {$k_i$};
\foreach \i in {0,...,8} {
    \pgfmathsetmacro{\x}{-4.1 + 0.34*\i}
    \node[base, fill=black!2] at (\x,0.25) {\pgfmathparse{\seq[\i]}\pgfmathresult};
}

\foreach \row/\start in {0/0,1/1,2/2,3/3} {
    \node[note, anchor=east] at (-4.5,-0.45-0.43*\row) {$s_{\row}$};
    \foreach \i in {0,...,5} {
        \pgfmathtruncatemacro{\idx}{\start+\i}
        \pgfmathsetmacro{\x}{-4.1 + 0.34*(\start+\i)}
        \pgfmathsetmacro{\y}{-0.45 - 0.43*\row}
        \node[hl] at (\x,\y) {\pgfmathparse{\seq[\idx]}\pgfmathresult};
    }
}

\draw[decorate,decoration={brace,amplitude=4pt}] (-4.27,-2.28) -- (-1.38,-2.28);
\node[note, green!50!black] at (-2.83,-2.62) {$(k-s+1)=4$ overlapping $s$-mers};

\node[box, minimum width=2.6cm] (hashes) at (2.35,-1.15) {$H$ bits per $s$-mer};
\node[box, minimum width=2.6cm] (present) at (2.35,-2.15) {present iff all bits set};
\draw[arrow, green!50!black] (-1.38,-1.47) -- (hashes.west);
\draw[arrow] (hashes.south) -- (present.north);

\end{tikzpicture}
    \end{minipage}%
}

\caption{Minimizer-guided sharding and $s$-mer verification, shown for $k=9$, $m=5$, $s=6$. Consecutive overlapping $k$-mers can share the same minimizer (here \texttt{GAAAC}) and therefore target the same shard. Within that shard, the findere scheme represents a $k$-mer through $(k-s+1)=4$ overlapping $s$-mers, requiring all corresponding bits to be set for a positive query.}
\label{fig:minimizer-findere}
\end{figure}

Fig. \ref{fig:findere-scheme} illustrates the Findere decomposition for a single $k$-mer.
Together, minimizer-guided sharding and Findere-based FPR suppression form the foundation of cuSBF's design: the expected 8-wide super-$k$-mer runs enable warp-level cooperation on the GPU (Section \ref{sec:insert-algorithm}), while the exponential FPR reduction provides strong accuracy even at modest shard sizes (Section \ref{sec:fpr}).

%\subsection{Minimizers(?)}

\subsection{Modern GPU Architectures}

 Processing on the GH200 (RTX PRO 6000 Blackwell) GPU is distributed across 132 (188) streaming multiprocessors (SMs) containing 4 vector units of 32 cores. Typically, one unit of work is assigned to one thread. When running a multi-threaded task, the GPU distributes small batches of threads (thread blocks) across SMs, which then schedule even smaller batches of 32 threads (warps) onto the 32-core vector units. Each SM has its own L1 cache (128 KB), which can be partially reconfigured to serve as fast user-programmable shared memory, allowing for fast data movement within a thread block. Similarly, each SM contains 256 KB worth of 32-bit registers, which can be cross-referenced by threads within a warp to exchange data via warp shuffling. Additionally, the GPU features a unified L2 cache (50 MB (128 MB) for GH200 (RTX PRO 6000)). When multiple neighboring threads within a warp load neighboring entries at the same time, the GPU performs a single, larger memory access instead, usually referred to as \emph{access coalescing}.

 Direct communication across SMs is restricted to communication through the GPU global main memory, which is based on GDDR or high-bandwidth on-chip memory (HBM).
 While the former typically achieve just above one terabyte per second of throughput, e.g., 1.8~TB/s on our utilized RTX PRO 6000, HBM is capable of sustaining several terabytes per second of throughput in the optimal case, e.g., 3.4~TB/s on GH200. This bandwidth, however, comes at the cost of relatively high access latency.
The minimum access granularity is 32B, referred to as a sector, with four sectors composing a 128B cache line.
%On HBM systems, or if ECC is enabled, the effective access granularity is 64B.
Accesses from threads on the same SM that target the same cache line or even sector can be merged into a single L2 request through temporal coalescing. Consequently, coalesced memory access patterns are critical for efficient use of GPU DRAM, as they minimize the number of high-latency memory transactions. Atomic updates benefit from the same coalescing mechanism as well, which is essential for supporting concurrent, lock-free insertions into the Bloom filter.

\section{Related Work}
\label{sec:related-work}

\textbf{Bloom filter variants.}
The standard Bloom filter \cite{bloom} and its cache-friendly blocked variant \cite{blocked-bloom} provide fast append-only AMQs. J\"unger et al. \cite{jünger2025optimizingbloomfiltersmodern} recently proposed optimized GPU Bloom filters using vectorization and thread cooperation. As
their implementation is not public, we use the \mbox{cuCollections} \cite{cuCollections} blocked Bloom filter as our high-performance GPU baseline.

\textbf{Sequence-aware CPU-based filters.}
The original Super Bloom filter \cite{superbloom} exploits minimizers to group adjacent $k$-mers and uses the Findere scheme \cite{findere} to reduce false positives through overlapping $s$-mer evidence.
In the authors' CPU benchmarks on streaming $k$-mer insert and query, it consistently outperforms classical and blocked Bloom filters as well as other widely used open-source implementations, with several-fold runtime gains at comparable memory budgets.
With Findere enabled it also achieves far lower false-positive rates than these baselines, and in a Rust reimplementation of BioBloom Tools it remains faster than filters built from standard or blocked Bloom layouts.
cuSBF keeps the same high-level AMQ semantics but redesigns the implementation around GPU execution, emphasizing register-resident bit masks, warp-level cooperation, and atomic reduction.

\textbf{Minimizer-based GPU $k$-mer processing.}
Minimum substring partitioning (MSP) groups consecutive $k$-mers sharing a minimizer into super-$k$-mers in order to reduce storage and partition workloads in $k$-mer counting \cite{minimizers}.
RapidGKC \cite{rapidgkc} accelerates the full MSP pipeline on GPUs, including a branch-divergence-minimizing signature rule for selecting common substrings, a parallel encoding scheme for variable-length super-$k$-mers, and pipelined CPU--GPU co-processing.
Gerbil \cite{gerbil} and GPU-KMC2 \cite{gpu-kmc2} likewise apply super-$k$-mer grouping for GPU-accelerated counting but do not accelerate the encoding or partitioning phases.
cuSBF shares the core insight that super-$k$-mers expose exploitable locality on GPUs, but applies it to AMQs rather than exact counting.
Instead of sorting super-$k$-mers into bins, cuSBF uses them to amortize shard loads and enable warp-level cooperation within a Bloom filter.
To our knowledge, cuSBF is the first GPU implementation to combine minimizer-based super-$k$-mer grouping with Findere-based Bloom filtering.

\textbf{Dynamic GPU filters.}
The Two-Choice filter and GPU Counting Quotient filter \cite{gpu-filters,gpu-qf} support dynamic operations on GPUs but rely on complex cooperative scheduling or shifting operations. Cuckoo-GPU \cite{cuckoo-gpu} demonstrates that dynamic AMQs can be made highly efficient, but cuSBF targets append-only sequence indexing where deletions are unnecessary and FPR is the primary constraint.

\textbf{Applications of sequence-based Bloom filters.}
Bloom filters are a core primitive in genomic sequence analysis pipelines, including large-scale dataset indexing and search in BIGSI \cite{bigsi}, Bloom-filter-based de Bruijn graph assembly in ABySS~2.0 \cite{abyss2}, metagenomic read classification in Ganon \cite{ganon}, short-read error correction in CUDA-EC \cite{shi2010parallel} and host-removal or contamination screening in BioBloom Tools \cite{biobloom}.
cuSBF targets the same workload family, but focuses on accelerating the underlying sequence-native insert and query path on GPUs.

\section{Design of cuSBF}
\label{sec:design}

\subsection{Library Design}
cuSBF is provided as a header‑only CUDA C++ library, allowing developers to integrate it without modifying their build system.
All tunable parameters ($k$, $s$, $m$, hash count $H$, CUDA block size, and the alphabet encoding) are exposed as compile‑time template arguments through a \texttt{Config}  structure. This enables the compiler to generate fully specialized kernels with static loop unrolling and constant propagation. The only run‑time input is the total filter size (in bits), which determines the number of shards.
%All configuration parameters ($k$, $s$, $m$, hash count $H$, CUDA block size, and the alphabet encoding) are exposed as compile-time template parameters through a \texttt{Config} structure, allowing the compiler to generate specialized code with static loop unrolling and constant propagation. The only runtime parameter is the total number of bits used by the filter, which determines the shard count.

The library supports three alphabet encodings out of the box: DNA (4 symbols, 2 bits), proteins (26 symbols, 5 bits), and DNA triplets (64 symbols, 6 bits, multi-byte).
While cuSBF does support non-DNA sequences, DNA is our primary target and the sole focus of this evaluation.
Larger alphabets remain functional, but they are a less favorable fit for this design because packing into 64-bit words leaves room for fewer symbols, which in turn forces smaller feasible parameter choices.
In particular, the smaller minimizer lengths that follow from large symbol widths substantially hurt false-positive rate, as the parameter sweep in Section~\ref{sec:parameter} shows.
cuSBF accepts several input forms that match common sequence-processing pipelines, including host-resident sequences, device-resident sequences, dense record batches, and streamed FASTA/FASTQ input with transparent gzip decompression.
Host-facing entry points synchronize before returning, while device-facing entry points leave stream ordering and synchronization under the caller's control.

\subsection{Data Layout}
The filter resides as a contiguous array of $2^P$ \emph{shards} in global memory. The shard count rounded up to the next power of two. This enables shard index computation with a single bitwise AND, avoiding a costly division.

Each shard is a 256‑bit (32‑byte) struct consisting of four 64‑bit unsigned‑integer words. This size matches the maximum single‑instruction load width on Blackwell‑class GPUs (Section \ref{sec:optimizations}).

The $H$ hash functions are distributed across the four word sectors: a hash index $i \in \{0, \dots, H-1\}$  maps to sector $s = i \bmod 4$. By concentrating a hash function’s bit updates within a single 64‑bit word, we minimize the atomic‑update footprint and enable independent word‑level operations.

%The filter's storage is a contiguous array of $2^P$ \emph{shards} allocated in GPU global memory. The shard count is rounded up to the next power of two, enabling shard index computation via fast bitwise AND instead of an expensive modulo.

%Each shard is a 32-byte aligned struct containing four 64-bit unsigned integer words. This 256-bit shard size was deliberately chosen to match the maximum single-instruction load width on Blackwell-class GPUs (Section \ref{sec:optimizations}).

%The $H$ hash functions are distributed across the four word \emph{sectors}: hash index $i \in \{0, \dots, H-1\}$ maps to sector $s = i \bmod 4$. This sectorized layout concentrates each hash function's bit updates within a single 64-bit word, reducing the atomic update footprint and enabling independent word-level operations.

\subsection{Shared Tile Preparation}

Insertion and query share the same front end. Rather than having each of $T$ threads
independently read $k$ overlapping symbols from global memory ($T \times k$ total reads), all threads cooperatively load a tile of $T + k - 1$ encoded symbols into shared memory.
Each thread loads symbols at stride $T$, achieving coalesced global memory access.
The tile also computes a tile-wide validity flag: if every symbol in the tile is valid, per-$k$-mer validity checks are skipped entirely.

Each active thread then packs one $k$-mer into a 64-bit integer using shift-and-OR operations.
For DNA this is possible for $k \leq 32$ because every base occupies two bits.
Larger alphabets use more bits per symbol and therefore support smaller maximum $k$ values.

\subsection{Insertion Algorithm}
\label{sec:insert-algorithm}

Each insertion thread processes one $k$-mer from the prepared tile (see Algorithm \ref{alg:insert}).
It first computes the minimizer by hashing every overlapping $m$-mer with a lightweight multiplicative hash and taking the minimum value.
The minimizer hash selects the target shard.
Then, for each of the $(k-s+1)$ overlapping $s$-mer windows, the thread hashes the $s$-mer and accumulates all corresponding bit positions into four 64-bit word masks held in registers.

\begin{figure}[t]
  \centering
  \begin{tikzpicture}[
      font=\sffamily\scriptsize,
      x=0.82cm,
      y=0.82cm,
      >=Stealth,
      thread/.style={
        rounded corners=1.5pt,
        draw=#1!75!black,
        fill=#1!12,
        very thick,
        minimum width=0.62cm,
        minimum height=0.48cm,
        inner sep=1pt,
        font=\sffamily\scriptsize\bfseries
      },
      thread/.default=black,
      inactive/.style={
        rounded corners=1.5pt,
        draw=black!25,
        fill=black!5,
        minimum width=0.62cm,
        minimum height=0.48cm,
        inner sep=1pt,
        font=\sffamily\scriptsize\bfseries,
        text=black!45
      },
      word/.style={
        rounded corners=1pt,
        draw=black!28,
        fill=white,
        minimum width=0.34cm,
        minimum height=0.22cm,
        inner sep=0pt,
        font=\sffamily\tiny
      },
      reducedword/.style={
        word,
        draw=#1!75!black,
        fill=#1!12,
        line width=0.7pt,
        font=\sffamily\tiny\bfseries
      },
      reducedword/.default=black,
      result/.style={
        rounded corners=1.5pt,
        draw=#1!75!black,
        fill=#1!12,
        very thick,
        minimum width=1.08cm,
        minimum height=0.40cm,
        inner sep=1pt,
        font=\sffamily\scriptsize\bfseries
      },
      result/.default=black,
      flow/.style={
        -{Stealth[length=1.7mm,width=1.25mm]},
        line width=0.55pt,
        draw=#1!78!black
      },
      flow/.default=black,
      faintflow/.style={
        -{Stealth[length=1.4mm,width=1.05mm]},
        line width=0.45pt,
        draw=#1!55!black,
        opacity=0.75
      },
      faintflow/.default=black,
      fan/.style={
        draw=#1!55!black,
        line width=0.45pt,
        opacity=0.75,
        line cap=round,
        line join=round
      },
      fan/.default=black,
      rowlabel/.style={
        anchor=east,
        font=\sffamily\scriptsize\bfseries,
        text=black!72
      },
      small/.style={
        font=\sffamily\tiny,
        text=black!65
      },
      callout/.style={
        font=\sffamily\tiny\bfseries,
        text=#1!75!black
      },
      callout/.default=black,
    ]

    \definecolor{runA}{RGB}{58,118,210}
    \definecolor{runB}{RGB}{43,154,87}
    \definecolor{runC}{RGB}{218,132,38}

    % Layout constants: one warp lane per x-coordinate.
    \def\xzero{0}
    \def\ythreads{0}
    \def\ymasks{-1.15}
    \def\yreduce{-2.50}
    \def\ywrite{-3.45}

    % Row labels.
    \node[rowlabel] at (-0.85,\ythreads) {lanes};
    \node[rowlabel] at (-0.85,\ymasks) {masks};
    \node[rowlabel] at (-0.85,\yreduce) {per-word OR};
    \node[rowlabel] at (-0.85,\ywrite) {writes};

    % Threads grouped by contiguous shard runs.
    \foreach \t in {0,1,2} {
      \node[thread=runA] (T\t) at (\t,\ythreads) {T\textsubscript{\t}};
    }
    \foreach \t in {3,4} {
      \node[thread=runB] (T\t) at (\t,\ythreads) {T\textsubscript{\t}};
    }
    \foreach \t in {5,6} {
      \node[thread=runC] (T\t) at (\t,\ythreads) {T\textsubscript{\t}};
    }
    \node[inactive] (T7) at (7,\ythreads) {T\textsubscript{7}};

    % Compact shard labels and run braces.
    \foreach \t/\shard/\col in {0/A/runA,1/A/runA,2/A/runA,3/B/runB,4/B/runB,5/C/runC,6/C/runC} {
      \node[small, text=\col!75!black] at (\t,\ythreads-0.46) {\shard};
    }
    \node[small, text=black!40] at (7,\ythreads-0.46) {sentinel};

    \draw[decorate, decoration={brace, amplitude=3pt}, runA!75!black, very thick]
    ($(T0.north west)+(0,0.34)$) -- ($(T2.north east)+(0,0.34)$)
    node[midway, above=3pt, callout=runA] {run A};
    \draw[decorate, decoration={brace, amplitude=3pt}, runB!75!black, very thick]
    ($(T3.north west)+(0,0.34)$) -- ($(T4.north east)+(0,0.34)$)
    node[midway, above=3pt, callout=runB] {run B};
    \draw[decorate, decoration={brace, amplitude=3pt}, runC!75!black, very thick]
    ($(T5.north west)+(0,0.34)$) -- ($(T6.north east)+(0,0.34)$)
    node[midway, above=3pt, callout=runC] {run C};

    % Per-lane four-word masks shown as a tiny vertical stack.
    \foreach \t in {0,1,2,3,4,5,6,7} {
      \node[word] at (\t-0.19,\ymasks+0.19) {$w_0$};
      \node[word] at (\t+0.19,\ymasks+0.06) {$w_1$};
      \node[word] at (\t-0.19,\ymasks-0.06) {$w_2$};
      \node[word] at (\t+0.19,\ymasks-0.19) {$w_3$};
    }

    % Reduced masks. Each run still produces four separate word reductions.
    \coordinate (RA) at (1,\yreduce);
    \coordinate (RB) at (3.5,\yreduce);
    \coordinate (RC) at (6,\yreduce);

    \foreach \x/\y/\label in {-0.19/0.19/$w_0$,0.19/0.06/$w_1$,-0.19/-0.06/$w_2$,0.19/-0.19/$w_3$} {
      \node[reducedword=runA] at ($(RA)+(\x,\y)$) {\label};
      \node[reducedword=runB] at ($(RB)+(\x,\y)$) {\label};
      \node[reducedword=runC] at ($(RC)+(\x,\y)$) {\label};
    }

    \coordinate (RAmerge) at ($(RA)+(0,0.54)$);
    \coordinate (RBmerge) at ($(RB)+(0,0.54)$);
    \coordinate (RCmerge) at ($(RC)+(0,0.54)$);
    \coordinate (RAin) at ($(RA)+(0,0.34)$);
    \coordinate (RBin) at ($(RB)+(0,0.34)$);
    \coordinate (RCin) at ($(RC)+(0,0.34)$);

    \foreach \t in {0,1,2} {
      \draw[fan=runA] (\t,\ymasks-0.35) .. controls (\t,\ymasks-0.82) and ($(RAmerge)+(0,0.16)$) .. (RAmerge);
    }
    \foreach \t in {3,4} {
      \draw[fan=runB] (\t,\ymasks-0.35) .. controls (\t,\ymasks-0.82) and ($(RBmerge)+(0,0.16)$) .. (RBmerge);
    }
    \foreach \t in {5,6} {
      \draw[fan=runC] (\t,\ymasks-0.35) .. controls (\t,\ymasks-0.82) and ($(RCmerge)+(0,0.16)$) .. (RCmerge);
    }

    \draw[faintflow=runA] (RAmerge) -- (RAin);
    \draw[faintflow=runB] (RBmerge) -- (RBin);
    \draw[faintflow=runC] (RCmerge) -- (RCin);

    % Only the run head performs the atomic write.
    \node[result=runA, minimum width=1.72cm] (WA) at (1,\ywrite) {$4\times$ atomicOr};
    \node[result=runB, minimum width=1.72cm] (WB) at (3.5,\ywrite) {$4\times$ atomicOr};
    \node[result=runC, minimum width=1.72cm] (WC) at (6,\ywrite) {$4\times$ atomicOr};

    \draw[flow=runA] ($(RA)+(0,-0.34)$) -- node[right, small, text=runA!75!black] {head T\textsubscript{0}} (WA);
    \draw[flow=runB] ($(RB)+(0,-0.34)$) -- node[right, small, text=runB!75!black] {head T\textsubscript{3}} (WB);
    \draw[flow=runC] ($(RC)+(0,-0.34)$) -- node[right, small, text=runC!75!black] {head T\textsubscript{5}} (WC);

    \node[small, align=center, text=black!70] at (3.5,-4.15)
    {Here, $7 \times 4 = 28$ atomic writes reduce to $3 \times 4 = 12$.};

  \end{tikzpicture}
  \caption{Segmented warp reduction. Consecutive threads sharing the same minimizer form a contiguous
    run (top). Within each run, \texttt{HeadSegmentedReduce} merges all four word masks via
    bitwise-OR using warp shuffles (middle). Only the run head issues \texttt{atomicOr} to the
  target shard in global memory (bottom).}
  \label{fig:segmented-reduction}
\end{figure}
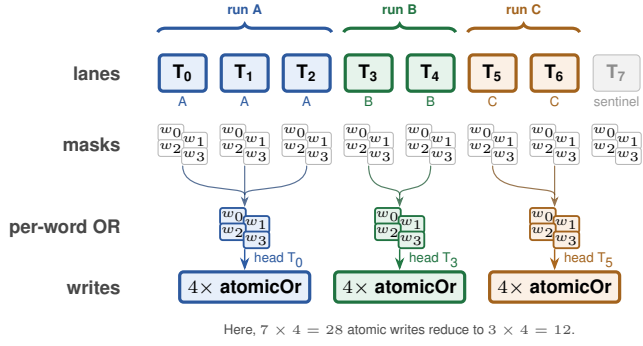

Contiguous threads that target the same shard form a run.
cuSBF uses CUB's \texttt{HeadSegmentedReduce} to merge the four word masks within each run using bitwise OR.
Only the run head issues \texttt{atomicOr} operations on the shard words. Fig. \ref{fig:segmented-reduction} illustrates this reduction.
This reduces the worst-case number of atomics from $T \times 4$ per CTA to approximately one per run per word.
Inactive threads, corresponding to invalid $k$-mers, receive a per-lane sentinel shard index, naturally splitting runs around them.

\begin{algorithm}
  [t]
  \caption{Parallel Insertion}
  \label{alg:insert}
  \footnotesize \Fn{InsertKmer(tile, threadIdx)}{
    \tcp*[l]{Cooperatively load encoded symbols into shared memory}
    LoadTile(tile, sequence, blockStart, blockKmers)\;

    \textnormal{active} $\gets$ ValidKmer(tile, threadIdx)\;
    \textnormal{packed} $\gets$ PackKmer(tile, threadIdx)\;

    \tcp*[l]{Minimizer: minimum hash across all m-mer windows}
    \textnormal{minHash} $\gets \infty$\;
    \For{$i \gets 0$ \KwTo $k-m$}{
      \textnormal{h} $\gets$ MinimizerHash(Subwindow(\textnormal{packed}, $i$, $m$))\;
      \If{$\textnormal{h}< \textnormal{minHash}$}{
        \textnormal{minHash} $\gets$ \textnormal{h}\;
      }
    }
    \textnormal{shardIdx} $\gets$ \textnormal{minHash} \&\ (\textnormal{NumShards}() - 1)\;

    \tcp*[l]{Accumulate s-mer hashes into 4 word masks}
    \textnormal{w0}, \textnormal{w1}, \textnormal{w2}, \textnormal{w3} $\gets 0, 0, 0, 0$\;

    \For{$i \gets 0$ \KwTo $k-s$}{
      \textnormal{sHash} $\gets$ SmerHash(Subwindow(\textnormal{packed}, $i$, $s$))\;
      HashToMasks(\textnormal{sHash}, \textnormal{\&w0}, \textnormal{\&w1}, \textnormal{\&w2}, \textnormal{\&w3})\;
    }

    \tcp*[l]{Detect contiguous run of threads targeting the same shard}
    \textnormal{prev} $\gets$ \textnormal{ShflUp}(\textnormal{shardIdx})\;
    \textnormal{runHead} $\gets (\textnormal{lane}= 0) \ \textbf{or}\ (\textnormal{prev}\ne \textnormal{shardIdx})$\;

    \If{$\neg\textnormal{active}$}{
      \textnormal{shardIdx} $\gets$ unique per-lane sentinel\;
    }

    \tcp*[l]{Merge masks within each run via segmented warp reduction}
    \textnormal{w0} $\gets$ HeadSegmentedReduce(\textnormal{w0}, \textnormal{runHead}, \textbf{OR})\; \textnormal{w1} $\gets$ HeadSegmentedReduce(\textnormal{w1}, \textnormal{runHead}, \textbf{OR})\; \textnormal{w2} $\gets$ HeadSegmentedReduce(\textnormal{w2}, \textnormal{runHead}, \textbf{OR})\; \textnormal{w3} $\gets$ HeadSegmentedReduce(\textnormal{w3}, \textnormal{runHead}, \textbf{OR})\;

    \tcp*[l]{Only the run head issues atomic writes}
    \If{\textnormal{runHead} \textbf{and} \textnormal{active}}{
      AtomicOr(\&\textnormal{shards}[\textnormal{shardIdx}].\textnormal{words}[0], \textnormal{w0})\;
      AtomicOr(\&\textnormal{shards}[\textnormal{shardIdx}].\textnormal{words}[1], \textnormal{w1})\;
      AtomicOr(\&\textnormal{shards}[\textnormal{shardIdx}].\textnormal{words}[2], \textnormal{w2})\;
      AtomicOr(\&\textnormal{shards}[\textnormal{shardIdx}].\textnormal{words}[3], \textnormal{w3})\;
    }
  }
\end{algorithm}

\subsection{Query Algorithm}

Query uses the same tile preparation and minimizer selection, but each thread processes $kStride = 4$ consecutive $k$-mers to amortize packing overhead.
The first $k$-mer is packed from scratch, and each subsequent $k$-mer is obtained by a single sliding-window operation: left-shift by the alphabet symbol width, OR in the new trailing symbol, and mask to $k$ symbols.
For $k=31$ and $kStride=4$, this reduces per-thread packing work from $4k=124$ loop iterations to $k+3=34$, saving $3(k-1)=90$ packing steps.

After computing a $k$-mer's target shard, threads within a warp call \texttt{\_\_match\_any\_sync} to identify lanes that need the same shard.
The lowest-numbered peer becomes the leader: it loads the entire 256-bit shard, then broadcasts the four words to peer lanes via \texttt{\_\_shfl\_sync}.
This replaces redundant global loads with register shuffles when adjacent $k$-mers share a minimizer.
The query then checks all $s$-mer windows and all $H$ hash functions, accumulating the result across all required bit tests. Algorithm \ref{alg:query} shows the overall process.

\begin{algorithm}
  [t]
  \caption{Parallel Query}
  \label{alg:query}
  \footnotesize \Fn{QueryKmers(tile, threadIdx)}{
    \tcp*[l]{Cooperatively load encoded symbols into shared memory}
    LoadTile(tile, sequence, blockStart, blockKmers)\;

    \textnormal{threadOffset} $\gets$ \textnormal{threadIdx} $\cdot$ \textnormal{kStride}\;

    \tcp*[l]{First k-mer packed from scratch}
    \textnormal{packed} $\gets$ PackKmer(\textnormal{tile}, \textnormal{threadOffset})\;

    \For{$s \gets 0$ \KwTo \textnormal{kStride} $- 1$}{
      \textnormal{localIdx} $\gets$ \textnormal{threadOffset} $+ s$\;

      \If{$s > 0$}{
        \tcp*[l]{Slide window: one shift-and-OR per subsequent k-mer}
        \textnormal{packed} $\gets$ AdvanceWindow(\textnormal{packed}, \mbox{\textnormal{tile}[\textnormal{localIdx} $+ k - 1$]})\;
      }

      \If{$\neg$ValidKmer(\textnormal{tile}, \textnormal{localIdx})}{
        \textnormal{output}[\textnormal{blockStartKmer} $+$ \textnormal{localIdx}] $\gets \textnormal{false}$\;
        \textbf{continue}\;
      }

      \tcp*[l]{Minimizer selects target shard}
      \textnormal{minHash} $\gets \infty$\;
      \For{$i \gets 0$ \KwTo $k-m$}{
        \textnormal{h} $\gets$ MinimizerHash(Subwindow(\textnormal{packed}, $i$, $m$))\;
        \If{$\textnormal{h}< \textnormal{minHash}$}{
          \textnormal{minHash} $\gets$ \textnormal{h}\;
        }
      }

      \textnormal{shardIdx} $\gets$ \textnormal{minHash} \&\ (\textnormal{NumShards}() - 1)\;

      \tcp*[l]{Warp-level shard sharing: one lane loads, all receive via shuffle}
      \textnormal{peers} $\gets$ MatchAny(\textnormal{shardIdx})\;
      \textnormal{leader} $\gets$ FirstLane(\textnormal{peers})\;

      \If{$\textnormal{lane}= \textnormal{leader}$}{
        Load256Bit(\textnormal{words}[0..3], \textnormal{shards}[\textnormal{shardIdx}])\;
      }

      \textnormal{words}[0..3] $\gets$ Shuffle(\textnormal{words}[0..3], \textnormal{leader}, \textnormal{peers})\;

      \tcp*[l]{Check all s-mer offsets and accumulate membership}
      \textnormal{present} $\gets$ Contains(\textnormal{packed}, \textnormal{words})\;

      \textnormal{output}[\textnormal{blockStartKmer} $+$ \textnormal{localIdx}] $\gets$ \textnormal{present}\;
    }
  }
  \BlankLine
  \footnotesize \Fn{Contains(packed, words)}{
    \textnormal{present} $\gets$ \textbf{true}\;

    \For{$i \gets 0$ \KwTo $k-s$}{
      \textnormal{sHash} $\gets$ SmerHash(Subwindow(\textnormal{packed}, $i$, $s$))\;

      \For{$h \gets 0$ \KwTo $H-1$}{
        \textnormal{sector} $\gets h \bmod 4$\;
        \textnormal{bitPos} $\gets$ BitAddress(\textnormal{sHash}, $h$)\;
        \textnormal{present} $\gets$ \textnormal{present} $\land$ BitIsSet(\textnormal{words}[\textnormal{sector}], \textnormal{bitPos})\;
      }
    }

    \textbf{return} \textnormal{present}\;
  }
\end{algorithm}

\subsection{Sequence-Native Design}

Unlike general-purpose filters that ingest individual integer keys, cuSBF operates directly on character sequences.
When processing FASTA/FASTQ input, consecutive records are concatenated with a designated \emph{separator} character (e.g., \texttt{'N'} for DNA) plus alignment padding to \texttt{symbolWidth} boundaries.
Any $k$-mer that spans a record boundary will contain the invalid separator and is naturally filtered out by the validity check.
This eliminates the need for record-boundary tracking inside GPU kernels, keeping the fast path simple and branch-free.

\subsection{Optimizations}
\label{sec:optimizations}

\textbf{Bit-slicing.}
Within each word sector, a hash value is mapped to a bit position in one of
two ways.
When $64/H \geq 6$, the 64-bit hash is divided into $H$ equal-width slices, and slice $i$ directly addresses a bit position for hash index $i$.
This avoids an extra multiplication for each hash function.
When $H$ is large and each slice is too narrow, cuSBF falls back to multiplying with an inlined per-hash-index golden-ratio-derived salt and extracting the upper bits.

\textbf{Role-specialized hashing.}
cuSBF uses different hash costs for different tasks.
Minimizer selection hashes each $m$-mer with a single multiplicative mix, which is sufficient for uniform shard selection and minimum comparison, while $s$-mer hashing for Bloom bit placement uses a stronger mix with an additional xorshift.
This reduces ALU work in the minimizer hot path without changing the higher-quality hashing used for filter updates and queries.

\textbf{Non-coherent vector loads.}
On Blackwell-class GPUs, query loads use \texttt{ld.global.nc.v4.u64} to fetch all four shard words in one instruction while bypassing L1. Older GPUs fall back to two 128-bit non-coherent loads via \texttt{ld.global.nc.v2.u64}.

\textbf{Compile-time specialization.}
All core loops over hash functions, $s$-mer windows, and minimizer windows are controlled by compile-time constants. This lets the compiler unroll hot loops and eliminate unused code paths for each configuration.

\textbf{Host FASTX staging.}
For FASTA/FASTQ inputs, cuSBF selects among stream, memory-mapped, and pipelined paths from file size, available host RAM, and a VRAM-derived staging budget (\texttt{fill\_fraction} of free GPU memory).
Uncompressed files that fit in memory can be \texttt{mmap}'d and parsed without an extra full-file read.
Multi-chunk workloads overlap host normalization, pinned staging, and H2D copies with GPU work on alternating CUDA streams.
When the entire normalized batch fits one chunk, each insert or query pass issues a single kernel launch.

\section{Evaluation}
\label{sec:evaluation}

\subsection{Experimental Setup}
The performance evaluation was conducted on three distinct hardware configurations.

\begin{itemize}
    \item \textbf{System A}: An AMD EPYC 7713P (64 cores) paired with an RTX PRO 6000 Blackwell GPU featuring 96 GB of GDDR7 memory (1.8 TB/s) running AlmaLinux 10.1 and CUDA 13.2.

    \item \textbf{System B}: A GH200 Grace Hopper system with 72 ARM Neoverse V2 cores and an H100 GPU featuring 96 GB of HBM3 (3.4 TB/s) running Ubuntu 24.04.4 and CUDA 13.2.

    \item \textbf{System C}: Intel\textsuperscript{\textregistered} Xeon\textsuperscript{\textregistered} W9-3595X CPU featuring 60 cores and 256 GB of DDR5 (300 GB/s) running AlmaLinux 10.1.
\end{itemize}

Systems A and B were selected to expose different architectural trade-offs.
While System B offers substantially higher memory bandwidth through HBM3 memory compared to GDDR7, System A provides roughly 50\% more CUDA cores.
This distinction helps differentiate compute-bound from memory-bound behavior.
System C, in contrast, has been chosen for its strong multi-core CPU performance and is used to evaluate the CPU reference implementation using 120 threads.

To provide a comprehensive comparison, we evaluate cuSBF against the following baselines:
\begin{itemize}
    \item \textbf{GPU Blocked Bloom filter (GBBF)}: NVIDIA's \mbox{cuCollections} Bloom filter \cite{cuCollections}, used as the high-performance append-only GPU baseline.

    \item \textbf{Cuckoo-GPU}: A modern GPU Cuckoo filter \cite{cuckoo-gpu}, included to compare against a high-throughput dynamic AMQ that supports insertions, queries, and deletions.

    \item \textbf{Bulk Two-Choice filter (TCF)}: A GPU-focused dynamic AMQ from McCoy et al. \cite{gpu-filters} that emphasizes locality through cooperative scheduling.

    \item \textbf{GPU Counting Quotient filter (GQF)}: The quotient-filter variant from the same study \cite{gpu-filters,gpu-qf}, included as a space-efficient dynamic baseline.

    \item \textbf{CPU Super Bloom}: The CPU reference implementation of Super Bloom \cite{superbloom}, used to separate algorithmic gains from GPU-specific acceleration.
\end{itemize}

All throughput comparisons use the same sequence-derived $k$-mer workloads and assign each filter a common nominal budget of 16 bits per inserted $k$-mer, rounded up to implementation-compatible capacities.
Unless noted otherwise, our sequence benchmarks use the \emph{C.~elegans} reference ($\approx$ 97 MiB) as the smaller, more cache-friendly workload and as the base dataset for the false-positive-rate experiment, and the human \emph{T2T-CHM13~v2.0} reference ($\approx$ 3 GiB) as a large, out-of-cache workload that stresses sustained global-memory behavior.

For the C.~elegans workload, cuSBF, GBBF, Cuckoo-GPU, TCF, and Super Bloom therefore use 256 MiB filters, while GQF uses the same nominal capacity but occupies 290.25 MiB because of its layout overhead.
For the human \mbox{T2T-CHM13~v2.0} workload, the corresponding sizes are 8 GiB for cuSBF, GBBF, Cuckoo-GPU, TCF, and Super Bloom, and 9.06 GiB for GQF.
Reported sizes count only the allocated filter object, scratch buffers and staging memory are excluded.
We also assume GPU-resident input, so host--device transfer time is not charged.

\subsection{Parameter Exploration}
\label{sec:parameter}

\begin{figure}[t]
  \centering
  \includegraphics[width=\linewidth]{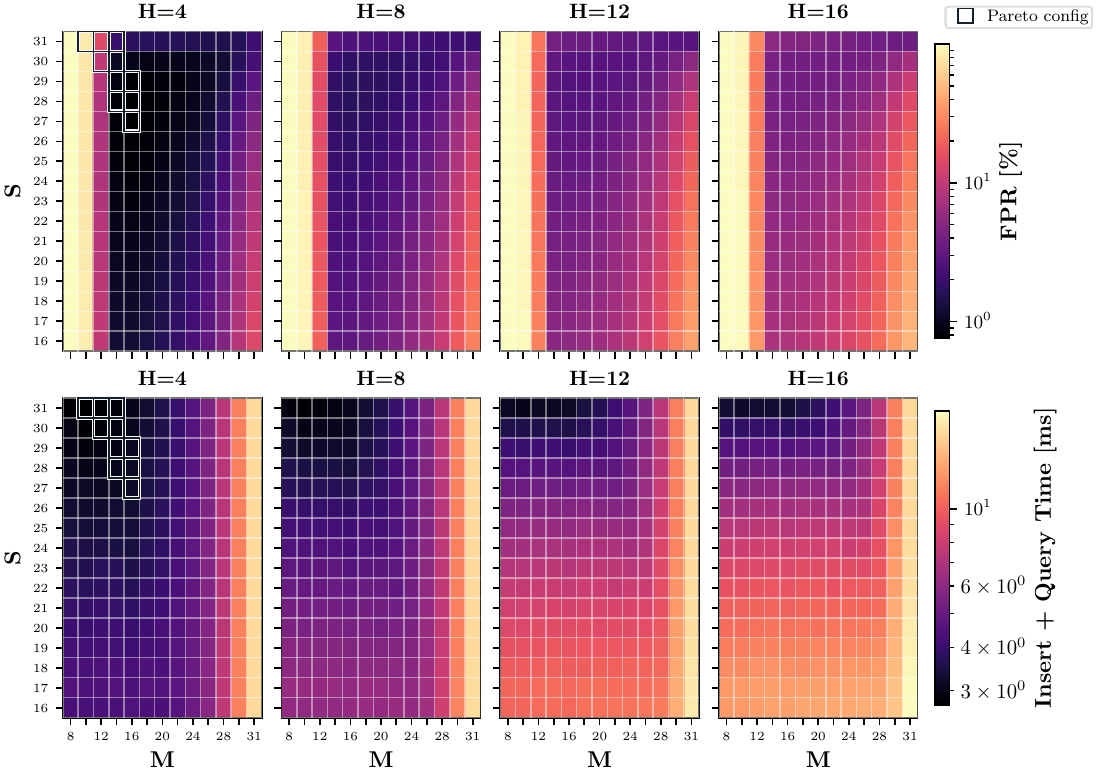}
  \caption{Parameter sweep summary for $k=31$ on System~A. The top row shows false-positive rate and the bottom row the combined insert+query time. Columns correspond to the hash count $H \in \{4,8,12,16\}$, while each heatmap sweeps the Findere length $s$ (vertical axis) and minimizer length $m$ (horizontal axis). Outlined cells are Pareto-optimal when minimizing false-positive rate and total runtime jointly.}
  \label{fig:param-sweep-heatmap}
\end{figure}

To isolate SBF trade-offs, we swept $(s,m,H)$ on the \emph{C.~elegans} workload at a 256 MiB memory budget (Fig.~\ref{fig:param-sweep-heatmap}).
All Pareto-optimal configurations lie at $H=4$, increasing $H$ raises runtime without improving the false-positive rate.
At our selected operating point of $(28,16,4)$, moving to $H=8$ increases runtime by 7.6\% and doubles the FPR, while $H=16$ raises runtime by 65.9\% and the FPR by $5.2\times$.

For fixed $(m,H)=(16,4)$, the Findere length $s$ is optimal around $s \approx 27$--29.
Increasing $s$ from 16 to 28 reduces total time by 29.1\% and the FPR by 37.4\% by reducing overlapping $s$-mers.
Beyond $s=28$, accuracy degrades because too few overlapping $s$-mers remain to suppress false positives; relative to $s=28$, the FPR is 25.6\% higher at $s=30$ and $2.29\times$ higher at $s=31$.

The minimizer length $m$ similarly peaks at $m=16$.
At $(s,H)=(28,4)$, increasing $m$ from 8 to 12 reduces the FPR by 91.7\%, and moving to 16 reduces it by another 90.8\%.
Beyond $m=16$, extra minimizer work and shorter super-$k$-mer runs degrade performance: at $m=18$, total time is 8.5\% higher and FPR is 3.5\% higher, and by $m=31$, insertion time is $7.7\times$ higher and FPR is $4.8\times$ higher.

Based on this frontier, we use $k=31$, $s=28$, $m=16$, and $H=4$ in the remaining benchmarks.
It sits at the low-FPR turning point of the Pareto set.
Compared to $(s,m,H)=(27,16,4)$ it retains essentially identical accuracy for 0.7\% lower total runtime, while compared to $(s,m,H)=(29,16,4)$ it pays only 2.3\% more time to reduce the FPR by 5.1\%.

\subsection{Throughput}
\label{sec:throughput}

\begin{figure*}[t]
  \centering
  \includegraphics[width=0.75\linewidth]{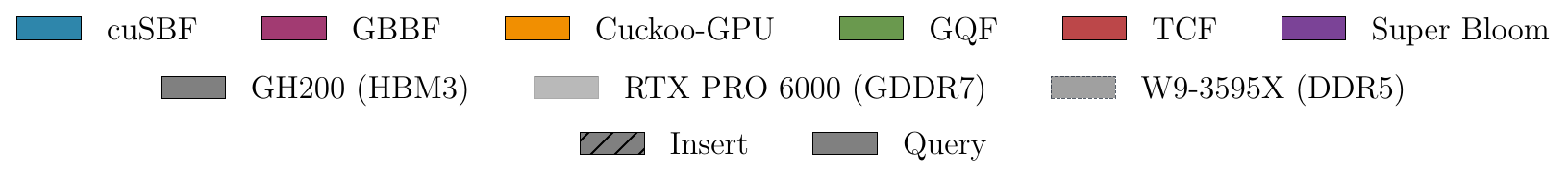}\par\medskip
  \begin{minipage}[t]{0.49\linewidth}
    \centering
    \includegraphics[width=\linewidth]{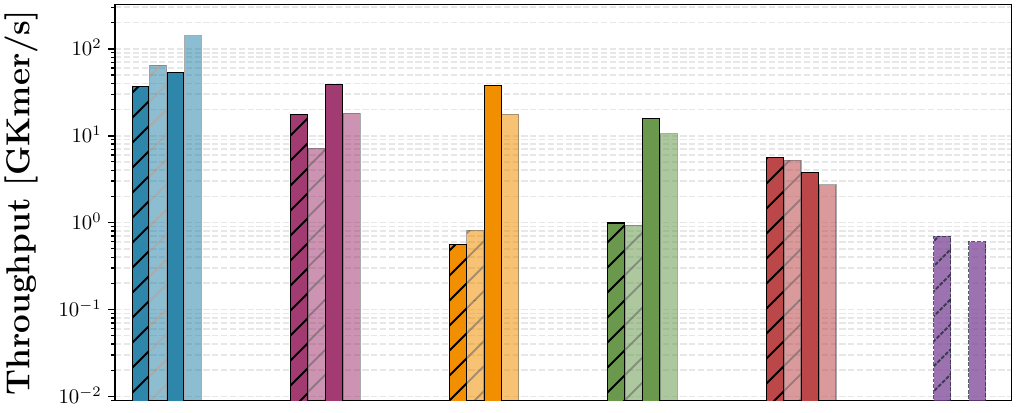}\\[2pt]
    \footnotesize (a) C.~elegans genome.
  \end{minipage}\hfill
  \begin{minipage}[t]{0.49\linewidth}
    \centering
    \includegraphics[width=\linewidth]{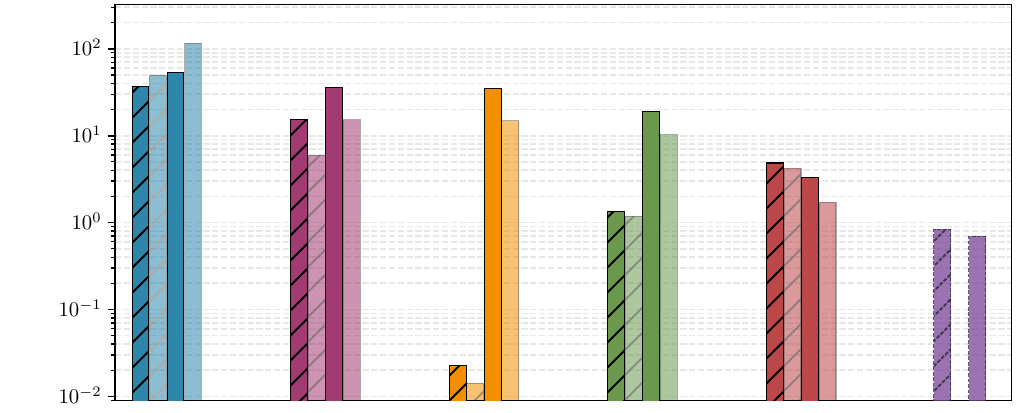}\\[2pt]
    \footnotesize (b) Human genome (T2T-CHM13).
  \end{minipage}
  \caption{Insert and query throughput on real genomic FASTA workloads across System~A (GDDR7), System~B (HBM3) and System~C (DDR5).}
  \label{fig:fastx-throughput}
\end{figure*}

Fig.~\ref{fig:fastx-throughput} shows that cuSBF remains the fastest method across both genomic workloads, though the extent of this advantage is heavily influenced by the available compute resources.
On System~A (GDDR7), cuSBF leads the \mbox{cuCollections} blocked Bloom baseline by about 9.1$\times$ for insertion and 7.7$\times$ for query on the smaller \emph{C.~elegans} filter, and by 8.2$\times$ and 7.6$\times$ on the larger human CHM13 filter.
Against the CPU Super Bloom implementation on System~C, the gap is even larger, reaching 92$\times$/234$\times$ on \emph{C.~elegans} and 59$\times$/165$\times$ on CHM13 for insert/query, respectively.

The relative picture changes on System~B (HBM3):
cuSBF still outperforms every baseline, but its advantage over the blocked Bloom filter narrows to 2.1$\times$ insertion and 1.37$\times$ query on \emph{C.~elegans}, and 2.36$\times$ and 1.49$\times$ on CHM13.
This is exactly the pattern that shows up in the speed-of-light analysis in Section \ref{sec:sol}: cuSBF is more compute-heavy, whereas \mbox{cuCollections} is closer to a pure bandwidth-bound baseline.
Moving from GDDR7 to HBM3 therefore helps \mbox{cuCollections} much more strongly than cuSBF.
\mbox{cuCollections} gains roughly 2.1--2.6$\times$ across these workloads, while cuSBF does not scale up and in fact runs slower on System~B, consistent with System~A offering substantially more CUDA cores despite the lower memory bandwidth.

Even so, cuSBF remains far ahead of the dynamic GPU AMQs on both GPU systems.
On System~A, it is about 79$\times$ faster than Cuckoo-GPU for insertion on \emph{C.~elegans} and over 3400$\times$ faster on CHM13, while still holding roughly 8$\times$ higher query throughput.
On System~B, the same comparison yields about 65$\times$ and 1600$\times$ insertion speedups and about 1.5$\times$ query speedups.
Relative to TCF and GQF, cuSBF's margins are also consistently large: on System~A it exceeds TCF by about 12$\times$ for insertion and over 50$\times$ for query, while on System~B it still maintains about 7$\times$ insertion and 15$\times$ query speedups.
These gaps show that even when cuSBF's own lead over blocked Bloom shrinks on HBM3, the sequence-aware design still translates into a clearly superior practical AMQ for sequence workloads.

Cuckoo-GPU deserves a separate discussion because its behavior is highly asymmetric.
Its query throughput stays close to the blocked Bloom baseline on both GPU systems, showing that read-only membership checks tolerate the duplicate-heavy genomic workload reasonably well.
Insertion, however, collapses on the human genome: relative to its own \emph{C.~elegans} throughput, Cuckoo-GPU drops by about 57$\times$ on System~A and 25$\times$ on System~B.
For genomic data this is unsurprising: DNA contains many duplicate $k$-mers, and duplicate insertions repeatedly target the same Ccuckoo buckets.
That amplifies contention and eviction churn, which is especially harmful for a write-heavy dynamic structure.
The smaller filter suffers from the same effect, but more traffic remains cache-resident and the penalty is less severe.
If the input would be deduplicated before insertion, Cuckoo-GPU would be placed much closer to the blocked Bloom baseline.

Overall, the throughput results show a clear architectural split.
cuSBF is not the most bandwidth-scalable design in the comparison.
Instead, it trades some peak memory throughput for much stronger end-to-end sequence throughput via minimizer-guided shard reuse, warp-level shard sharing, and segmented atomic reduction.
That trade-off remains favorable on both GPU platforms and leaves the CPU implementation far behind.

\subsection{False Positive Rate}
\label{sec:fpr}

\begin{figure}[t]
  \centering
  \includegraphics[width=\linewidth]{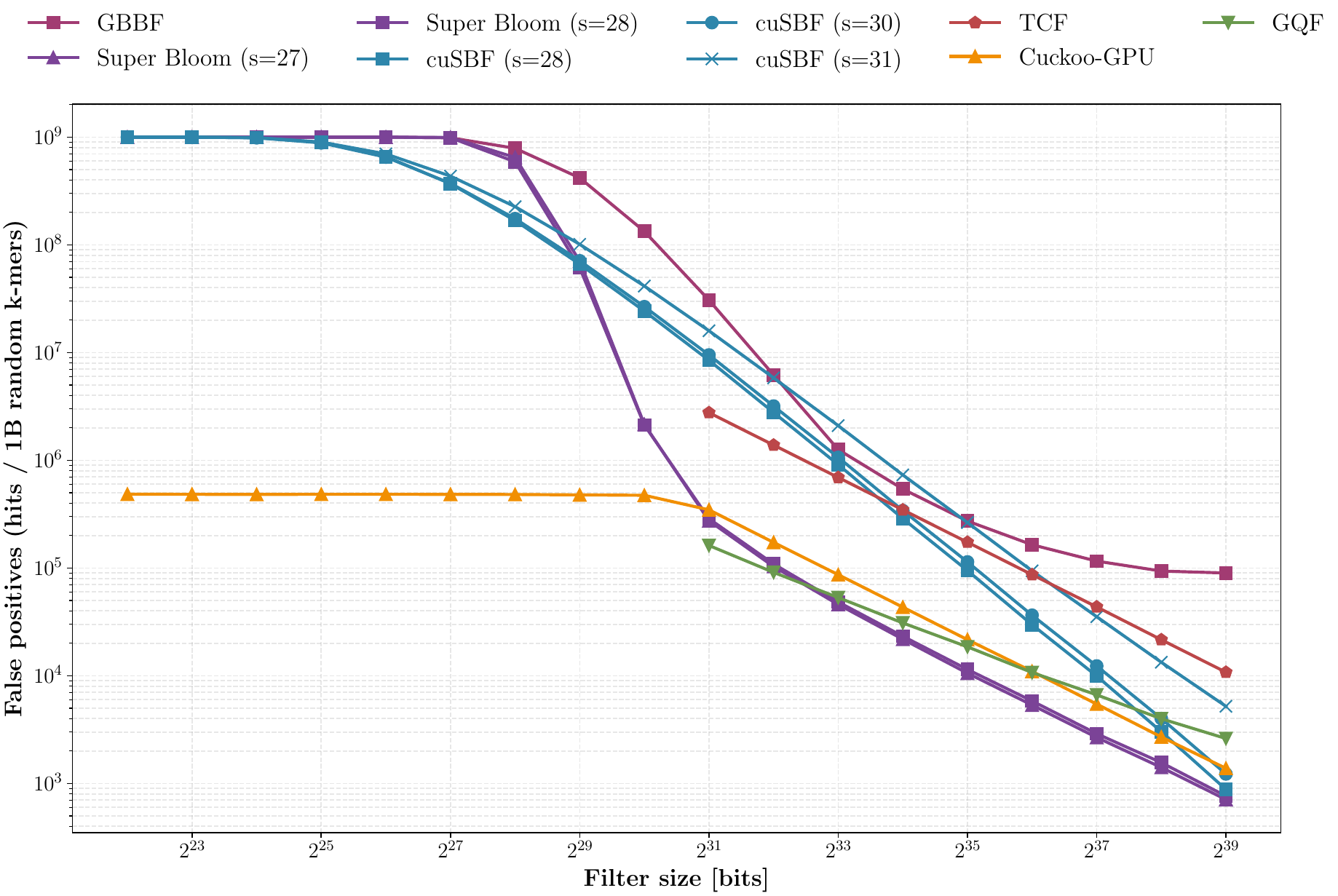}
  \caption{False-positive hits among $10^9$ random 31-mers after inserting the C.~elegans reference, plotted against allocated filter size.}
  \label{fig:fpr-fastx}
\end{figure}

Fig.~\ref{fig:fpr-fastx} plots the number of false positives among $10^9$ random 31-mers after inserting the \emph{C.~elegans} reference while sweeping the total filter size from $2^{22}$ to $2^{39}$ bits.
Among the cuSBF variants, the expected ordering from Eq. \eqref{eq:findere-fpr} is visible immediately: $s=28$ yields the lowest FPR, $s=30$ is consistently worse, and $s=31$ (which removes the Findere overlap entirely) is the worst throughout.
This is the cost of enlarging $s$: the per-$s$-mer test becomes more selective, but the number of overlapping $s$-mers per 31-mer drops from 4 to 2 to 1, weakening the multiplicative suppression of random-alien false positives.

The primary accuracy target for us is the \mbox{cuCollections} blocked Bloom baseline.
cuSBF with $s=28$ outperforms that baseline at every shared budget, delivering 3.6$\times$ lower FPR at $2^{31}$ bits, 2.9$\times$ lower FPR at $2^{35}$ bits, and just over two orders-of-magnitude lower FPR at $2^{39}$ bits.
The gap widens as the filters grow because the blocked Bloom curve begins to flatten at large capacities.
This is consistent with cuco's default single-hash blocked policy, which uses one 64-bit hash both to choose the block and to derive the in-block bit pattern.
Once the filter is extremely large, adding more memory mostly creates more blocks rather than proportionally strengthening the evidence checked inside each block.

TCF and GQF only appear from $2^{31}$ bits onward, but for different reasons: below that point TCF's additional scratch buffers no longer fit alongside the filter at the fixed 95\% load factor, whereas GQF's allocated filter object is already too large because of its layout overhead.
Of those two, GQF is clearly better: it stays roughly 8$\times$ below TCF throughout the shared range, beats Cuckoo-GPU at smaller and mid-sized budgets, and only falls behind once the sweep reaches the very largest filters.
Cuckoo-GPU is strong across the entire range, dropping by about 251$\times$ from $2^{31}$ to $2^{39}$ bits, and it stays below cuSBF until the very largest point: it is about 24$\times$ lower than cuSBF at $2^{31}$ bits, but by $2^{39}$ bits cuSBF is about 1.6$\times$ lower.
Its slope is nevertheless shallower than the Bloom-like filters, which is consistent with the benchmark configuration keeping a fixed 16-bit fingerprint and 16-slot bucket layout while scaling capacity: additional memory lowers occupancy, but it does not widen the fingerprints that ultimately control accidental matches.

The CPU Super Bloom reference retains the best FPR among the SBF-family variants, remaining far below cuSBF at moderate budgets and still slightly ahead at the largest one: relative to cuSBF with $s=28$, it is about 30$\times$ lower at $2^{31}$ bits, 8.3$\times$ lower at $2^{35}$ bits, and 1.15$\times$ lower at $2^{39}$ bits.
This is where our GPU-oriented simplifications show up most clearly: the CPU reference uses larger 512-bit blocks, and its per-$s$-mer hash applies a stronger SplitMix-style 64-bit mix.
By contrast, cuSBF sectorizes each 256-bit shard into four 64-bit words, so each hash index addresses only one 64-bit sector rather than the whole shard.
That reduces bit dispersion, but avoids the runtime mask-selection branching that would otherwise be far too costly on the GPU.
That trade-off is intentional.
In the intended deployment regime, throughput is the primary constraint as long as the FPR remains low enough to make the filter useful.
Under that criterion cuSBF lands in the right part of the design space: it consistently beats the GPU blocked Bloom baseline on FPR, while Section~\ref{sec:throughput} showed that it surpasses every evaluated alternative by a wide margin in throughput.

\subsection{Speed-of-Light Analysis}
\label{sec:sol}

\begin{figure}
    \centering
    \includegraphics[width=\linewidth]{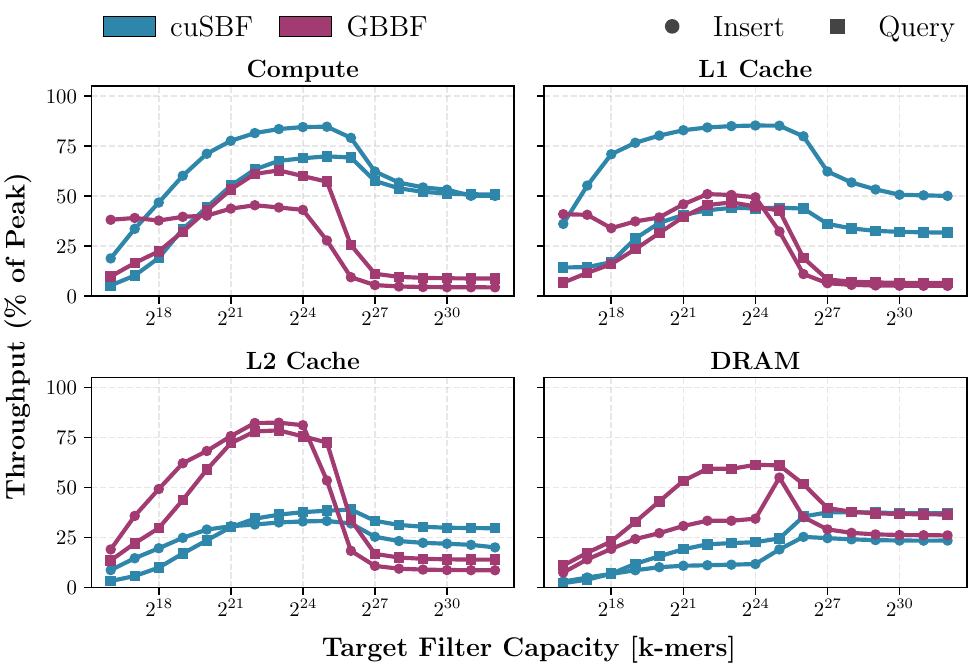}
    \caption{Speed-of-Light throughput metrics for cuSBF and cuco blocked Bloom on System~A across varying target capacities, fixed at 16 bits per inserted $k$-mer and a constant 95\% load factor.
    Compute, L1, L2, and DRAM throughput are reported as percentages of peak sustained performance.}
    \label{fig:sol}
\end{figure}

Fig.~\ref{fig:sol} compares cuSBF against the \mbox{cuCollections} blocked Bloom filter baseline on System~A with a fixed memory budget of 16 bits per inserted $k$-mer and a constant 95\% load factor.
cuSBF reaches about 85\% of peak SM throughput for insertion and about 70\% for query while keeping L2 and DRAM utilization comparatively modest over a broad range of capacities.
This indicates that the optimized cuSBF kernels are primarily limited by on-chip work (packing, minimizer selection, hashing) rather than by off-chip bandwidth alone.
Only once the filter grows beyond the size of the L2 cache does DRAM pressure rise noticeably, yet cuSBF still sustains about 50\% SM throughput at the largest tested capacities.

In contrast, cuco reaches high L2 and DRAM utilization much earlier, but its SM throughput collapses once the filter outgrows the cache-resident regime.
For insertion, cuco peaks at about 82\% L2 throughput but falls below 10\% SM throughput at the largest capacities.
For query, DRAM throughput rises to about 61\% while SM throughput drops to below 9\%.
This is consistent with cuSBF's minimizer-guided shard reuse and segmented warp reductions: they keep more work local to registers and shared memory, reducing the cost of repeated random global memory accesses when the filter no longer fits comfortably in cache.

\subsection{Host-to-Device Transfer Overhead}
\label{sec:host-overhead}

Section~\ref{sec:throughput} assumes GPU-resident input, which matches a pipeline stage but not standalone use from host-resident FASTX files, where H2D transfer and synchronization can dominate.
This question is especially relevant on a new class of tightly coupled heterogeneous systems, including GH200, GB200 and GB300, Vera Rubin, and compact platforms such as DGX Spark and RTX Spark, which pair CPUs and GPUs behind cache-coherent chip-to-chip links rather than a conventional PCIe root complex.
On GH200, the NVLink-C2C interconnect advertises up to 450~GB/s per direction~\cite{schieffer2024harnessingintegratedcpugpumemory}, roughly 7$\times$ higher than PCIe Gen5.
Because Section~\ref{sec:throughput} nevertheless favored the RTX PRO~6000 once data was already on the device, we test whether that faster link wins when sequence bytes remain on the host.

\begin{figure}[t]
  \centering
  \includegraphics[width=\linewidth]{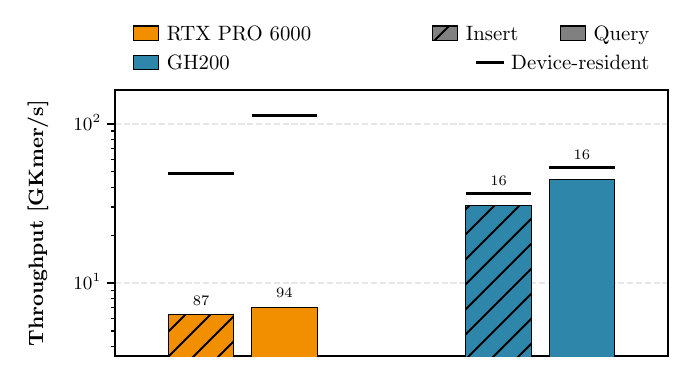}
  \caption{cuSBF insert and query throughput on the human  \emph{T2T-CHM13 v2.0} reference at the main benchmark configuration.
    Bars report throughput with host-resident sequence input, horizontal markers show device-resident throughput on the same system.
  Percentages quantify the throughput lost to host--device transfer relative to that device-resident rate.}
  \label{fig:host-overhead}
\end{figure}

Fig.~\ref{fig:host-overhead} compares host-sequence insert and query (excluding FASTX parsing) against device-resident async kernels on  \emph{T2T-CHM13 v2.0} at the main configuration.
On System~A, host insert and query reach only 6.3 and 7.1~GKmer/s, 87--94\% below the corresponding device-resident throughput of 48.5 and 112~GKmer/s.
On System~B, transfer overhead is smaller (17\% on insert, 16\% on query) and reverses the platform ranking: despite slower device-resident kernels, GH200 delivers 4.8$\times$ and 6.3$\times$ higher host-sequence throughput than the RTX PRO~6000.

\section{Conclusion}
\label{sec:conclusion}

We have introduced cuSBF, the first GPU‑native implementation of the Super Bloom filter that is tailored to genomic $k$-mer workloads. By combining minimizer‑guided sharding with a GPU‑oriented data layout (256‑bit sectorized shards, cooperative shared‑memory tiling, warp‑level shard sharing, and segmented warp reductions), cuSBF converts the locality inherent in super‑$k$‑mers into high‑throughput, low‑contention execution on modern accelerators.

%cuSBF brings the Super Bloom filter to modern GPUs as a sequence-native, header-only CUDA library for genomic $k$-mer workloads. Its design combines minimizer-guided sharding with a GPU-oriented implementation based on sectorized 256-bit shards, cooperative shared-memory tiling, warp-level shard sharing, and segmented warp reductions. Together, these choices turn the locality exposed by super-$k$-mers into a practical throughput advantage on modern accelerators.

Across the evaluated genomic workloads, cuSBF consistently delivers the highest end-to-end throughput among all tested sequence-capable baselines.
It maintains this advantage across both the discrete GDDR7-based RTX PRO 6000 and the HBM3-based GH200, even though the latter favors more purely bandwidth-bound designs.
At the same time, cuSBF does not trade away accuracy to obtain this speed.
The parameter sweep identifies $k=31$, $s=28$, $m=16$, and $H=4$ as the best overall operating point, and the resulting configuration consistently achieves lower false-positive rates than the GPU blocked Bloom baseline at comparable memory budgets.

The speed-of-light analysis clarifies why this trade-off works: cuSBF is not the most bandwidth-scalable filter in the comparison.
Instead, it deliberately spends more on-chip work to reduce redundant global-memory traffic and to exploit the structure of overlapping sequence windows.
For the intended deployment setting of GPU-native genomic pipelines, this proves to be the better design point.

At the same time, the current design has clear limitations that point to future work.
Although cuSBF supports arbitrary alphabets, larger alphabets consume more bits per symbol and therefore force smaller feasible minimizer lengths within the current 64-bit packing scheme, which in turn raises the false-positive rate.
One promising direction is to explore wider packed representations, such as multiword or SIMD/vector-assisted encodings, to support larger alphabets without giving up too much performance.
For standalone use over host-resident sequence data, Section~\ref{sec:host-overhead} further shows that such integrated platforms can outweigh faster discrete GPUs once data movement is included.

Overall, our results show that minimizer-aware AMQs can be mapped efficiently to GPUs and can substantially outperform generic GPU filters on real sequence data, but that the best future designs will likely need to co-optimize both kernel execution and data movement for the target system.

\balance
\bibliographystyle{IEEEtran}
\bibliography{bib-refs}
%\bibliography{IEEEabrv,bib-refs}

\end{document}